\documentclass[preprintnumbers,amsmath,amssymb,pra,showkeys]{revtex4-1}
\usepackage{color}
\usepackage{graphicx}

\def\bravert{\egroup\,\vrule\,\bgroup}

{\catcode`\|=\active
  \gdef\Twoint#1{\left(\mathcode`\|"8000\let|\bravert {#1}\right)}}
{\catcode`\|=\active
  \gdef\Braket#1{\left<\mathcode`\|"8000\let|\bravert {#1}\right>}}

\newcommand{\beq}{\begin{equation}}
\newcommand{\eeq}{\end{equation}}
\newcommand{\beqa}{\begin{eqnarray}}
\newcommand{\eeqa}{\end{eqnarray}}
\newcommand{\bea}{\begin{array}}
\newcommand{\eea}{\end{array}}

\newcommand{\bef}{\begin{figure}}
\newcommand{\ef}{\end{figure}}
\newcommand{\bc}{\begin{center}}
\newcommand{\ec}{\end{center}}
\newcommand{\bt}{\begin{table}}
\newcommand{\et}{\end{table}}
\newcommand{\btb}{\begin{tabular}}
\newcommand{\etb}{\end{tabular}}

\def\etal{\mbox{et al.}}

\begin{document}
\title {\itshape{What do we approximate and what are the consequences in perturbation theory?}}

\author{Lasse Kragh S{\o}rensen$^{a}$ \vspace{6pt},
        Roland Lindh$^{a,b}$\vspace{6pt} and
        Marcus Lundberg$^{a}$\vspace{6pt} \\
        $^{a}${\em{Department of Chemistry - {\AA}ngstr{\"o}m Laboratory,
                   Uppsala University,
                   S-75105 Uppsala,
                   Sweden}};
        $^{b}${\em{Uppsala Center of Computational Chemistry - UC$_3$,
                   Uppsala University,
                   S-75105 Uppsala,
                   Sweden}}\vspace{6pt}\received{v1.0 submitted july 2016}}

\email{lasse.kraghsorensen@kemi.uu.se}
\bigskip


\begin{abstract}
 
We present a discussion of the consequences in perturbation theory
when an exact eigenfunctions and eigenvalues to to the zeroth order Hamiltonian $H_0$
cannot be found. Since the usual approximations such as projecting
the wavefunction on to a finite basis set and restricting the
particle interaction is a way of constructing an approximate
zeroth order Hamiltonian $H_0'$ 
we will here argue that the exact eigenfunctions and eigenvalues
are always found for $H_0'$. We will show that as long as the perturbative
expansion does not depend on any intrinsic properties of $H_0$
but only on knowing the exact eigenfunctions and eigenvalues
then any perturbative statement, such as origin independence
intensities, will be true for any $H_0'$ provided that
$H_0'$ has a spectrum. 
We will use this to show that the origin independence for
the intensities is trivially fulfilled in the velocity gauge
but also can be fulfilled exactly in the length gauge if an 
appropriate $H_0$ is chosen.
Finally a small numerically demonstration 
of the origin dependence of the terms for the second-order
intensities in both the length and velocity
gauge is undertaking to numerically illustrate
the theoretical statements.

\end{abstract}

\maketitle

\section{Introduction}
\label{SEC:intro}
In perturbation theory the effect of a perturbation
is usually derived assuming 
that the exact eigenfunctions
and eigenvalues for the zeroth order Hamiltonian $\hat H_0$ are known \cite{lowdin_2,lowdin_4}.
For systems like a particle in a box, the harmonic
oscillator and other systems which can be solved
algebraically the exact eigenfunctions and eigenvalues
can of course be obtained, however, for most
applications of perturbation theory the exact
eigenfunctions and eigenvalues of $\hat H_0$ are not known.
Examples of this, related to electronic structure theory,
is the inclusion of an external electromagnetic
field that perturbs an atom or molecule, since it is
here assumed that the exact time-independent solution
of the atom or molecule is known, or even describing
the electron correlation with perturbation theory,
such as M{\o}ller-Plesset perturbation theory \cite{mppt},
since the SCF equations are solved in a finite basis set.
When the exact eigenfunctions $\hat H_0$ cannot be
found then in most textbooks it is stated that an approximate
wavefunction to $\hat H_0$ is found. The consequences
of not having the exact eigenfunctions and eigenvalues of $\hat H_0$
are rarely discussed if at all \cite{shavitt_2002}.

When the focus is on $\hat H_0$ the aim is to construct
a better $\hat H_0$ \cite{epstein,fink} and not on whether or not 
the exact solution to the given $\hat H_0$ can be found. 
Usually the the focus in perturbation theory has been on
the development of new types of perturbation expansions \cite{van_vleck,mppt,nevpt},
their relations \cite{shavitt_qdpt},
the convergence \cite{schucan} or lack thereof \cite{mpdiv,rajat,jompf},
bound for the energies \cite{lowdin_bounds}, eliminating of intruder states \cite{ih_malrieu,hoffmann},
conceptional developments of effective Hamiltonians \cite{freed,freed2,pradines},
multiconfigurationel \cite{andersson_caspt2_1,andersson_caspt2_2} 
or degenerate perturbation theory \cite{brandow,hose,klein_1974} just
to mention a few of the many developments that has
been on perturbation theory over many years. For
a more detailed historical account of the development
of perturbation theory we refer to \cite{lowdin}.

We will here show that the exact eigenfunctions and 
eigenvalues are always found but that these need not be
for the exact $\hat H_0$ but for some approximate or effective
zeroth order Hamiltonian $\hat H_0'$. Always having
the exact solution to $\hat H_0'$ means that any
perturbation statement will always be true for any choice
of basis set and level of correlation provided
that the perturbation statement is based on a
perturbation expansion which only require
that $\hat H_0'$ has a spectrum and not
on some intrinsic properties of $\hat H_0$.
An example of a perturbation statement is the
origin independence of higher order intensities \cite{bernadotte2012origin},
where an external electromagnetic field is applied
to a molecular system, and the perturbation treatment
is performed using Fermi's golden rule.

We will exploit the simple observation that
the exact eigenfunctions and eigenvalues for $\hat H_0'$
is always known to
show that the origin independence of higher order 
intensities \cite{bernadotte2012origin} always
hold in the velocity gauge but not in length gauge.
It will be shown that the problems in the length gauge
stems from not having the exact same commutation relations
for $\hat H_0$ and $\hat H_0'$ when transforming from
the velocity to the length gauge.
These findings will be backed by some numerical 
examples of exact and approximate origin dependence 
for certain electric and magnetic contributions to the
origin independent intensities \cite{bernadotte2012origin} for [FeCl$_4$]$^{1-}$
in the velocity and length gauge \cite{oiqm}.

\section{Theory}
\label{SEC:theo}

In the first two parts of this section we will discuss 
perturbation theory, with a particular focus on 
how the the zeroth order Hamiltonian $\hat H_0$
is constructed and what kind of consequences this has for
the perturbation expansion. We will here show
that the $\hat H_0$ usually assumed used
is in fact approximated by $\hat H_0'$
and as a consequence the exact eigenfunctions and eigenvalues
of the used zeroth order Hamiltonian $\hat H_0'$
is trivially found. We will here give an example
from Configuration-Interaction (CI) theory \cite{hyll_ci} on how
a series of approximate Hamiltonians can be constructed
from the exact solution.

Thereafter we will show that the error
in approximate calculations when
transforming from the velocity gauge to the
length gauge stems from the assumed non-exact commutation relation between
$\hat H_0'$ and $r$ and not from non-exact eigenfunctions
of $\hat H_0$.
Finally we will use the findings from the
construction of approximate Hamiltonians to
to show that the so-called quadrupole intensities,
recently derived by Bernadotte \etal \cite{bernadotte2012origin}, 
will be origin independent in the velocity gauge
irrespectively of the choice of basis set and
level of correlation. Here we will
repeat the equations essential to show
origin independence for self consistency, illustrate
where $\hat H_0'$ enters and the difference between
the length and velocity gauge and otherwise refer to 
the excellent paper by Bernadotte \etal \cite{bernadotte2012origin}
for complete derivations.

\subsection{Perturbation theory}
\label{sec:pert}

In perturbation theory the Hamiltonian $\hat H$ is divided into
a zeroth order Hamiltonian $\hat H_0$ and a perturbation $\hat U$
\beq
\label{hexact}
\hat H = \hat H_0 + \hat U
\eeq
where it is assumed that the exact eigenfunctions and eigenvalues for $\hat H_0$
are known and that the effect of $\hat U$ in some sense is 
sufficiently small so that the eigenfunctions of $\hat H$
can be expanded in the eigenfunctions of $\hat H_0$. 
The perturbation $\hat U$ is, however, independent of $\hat H_0$
so an alternative Hamiltonian $\hat H'$ with the
same perturbation $\hat U$
\beq
\label{hprime}
\hat H' = \hat H_0' + \hat U,
\eeq
where again it is assumed that the exact eigenfunctions and eigenvalues for $\hat H_0'$
is known, is also acceptable. Whether $\hat H_0$ or $\hat H_0'$ is used it will 
later be shown that the exact
same derivation for the oscillator strengths, in the velocity
representation, in Sec. \ref{perturb} could be performed
and exactly the same conclusion with respect to the origin
independence would be reached. 
In fact any conclusion reached for a perturbation expansion 
will always be true of any choice of $\hat H_0$ or $\hat H_0'$,
which only require knowledge of the exact eigenfunctions
and eigenvalues of $\hat H_0$, provided that $\hat H_0$ 
and $\hat H_0'$ has a spectrum. If, however, 
the perturbation treatment depend on some intrinsic
property of $\hat H_0$ the perturbation expansions
will then only be identical for another $\hat H_0'$
with the same intrinsic properties.
The intrinsic property of $\hat H_0$ could be some special commutation
relations with $\hat U$ that would simplify the
perturbation expansion or give some special conclusion.
We will here limit ourselves to perturbation expansions
which does not depend on any intrinsic properties of
$\hat H_0$ and hence the choice of
$\hat H_0$, and consequently $\hat H$,
can be choosen independently of $\hat U$.
One is in fact free to choose almost anything as 
$\hat H_0$ and $\hat U$,
even to include some fictitious interaction,
\beq
\label{fic}
\hat H = \hat H - \hat H_{fic} + \hat H_{fic} = \hat H_0 + \hat H_{fic} = \hat H_0 + \hat U
\eeq
where $\hat H_{fic}$ is some fictitious interaction.
If the exact eigenfunctions for $\hat H_0$ in Eq. \ref{fic}
can be found, make sense and give a convergent perturbation
series then this can be a practical way of solving the
eigenvalue problem for $\hat H$.

While it may seem strange to introduce some fictitious
interaction the well
known M{\o}ller-Plesset perturbation theory \cite{mppt} where
the perturbation operator $\hat \Phi$, known as the fluctuation operator,
\beq
\hat \Phi = \hat H - \hat f - h_{nuc}
\eeq
has the artificial mean-field description from Hartree-Fock $\hat f$
subtracted can be formulated as such.

\subsection{Approximations and exact eigenfunctions}
\label{approx}

In all perturbation calculations $\hat H_0$ is in 
some way approximated
except for those where an algebraic solution is known
like the harmonic oscillator, particle in a box et cetera.
The two major approximation usually performed in 
electronic struture theory is the projection 
of the wavefunction onto a finite basis
and the second in the interaction between
particles like truncating the CI hierarchy.
These approximations are usually thought of
as approximations in the wavefunction for the 
exact Hamiltonian but they
are in fact a way of creating an approximate or effective zeroth order
Hamiltonian $\hat H_0 '$ which is solved
exactly
\beq
\label{div}
\hat H_0 = \hat H_0 - \hat H_0' + \hat H_0' = \hat H_0' + \hat H_{rest}
\eeq
where the remaining effects from the finite basis and
incomplete correlation treatment is incorporated in $\hat H_{rest}$.
It may not be directly possible to write down $\hat H_{rest}$
for a specific system in a closed form, 
the division of the Hamiltonian in Eq. \ref{div}
is, however, still allowed.
In practice it is therefore the approximate zeroth order Hamiltonian $\hat H_0'$
that will be solved and not $\hat H_0$. Hence it is therefore
not the exact Hamiltonian $\hat H$ in Eq. \ref{hexact}
that is being solved but the alternative approximate or effective 
Hamiltonian $\hat H'$ in Eq. \ref{hprime} when
a perturbation is applied to the system.
Any perturbative derivation and conclusions
should therefore be based on $\hat H_0'$
and not $\hat H_0$. 

While the perturbation $\hat U$
is written as the same in Eqs. \ref{hexact} and \ref{hprime}
the effect of $\hat U$ will be affected by
the choice of $\hat H_0'$ and hence the
result of the perturbation will differ. 
We here note that what is called the exact 
Hamiltonian here is in fact arbitrary which is
in line with all current theories in physics where
any Hamiltonian is an effective theory dependent 
Hamiltonian. $\hat H_0'$ in Eq. \ref{div} therefore contain
the approximate or fictive interaction introduced in Eq. \ref{fic}.

The approximate Hamiltonian $\hat H_0'$ can then be
perturbed with $\hat U$ as shown in Eq. \ref{hprime}
in order to find a perturbative solution to $\hat H'$.
Since the exact eigenfunctions to $\hat H_0'$ will always be
found by construction,
disregarding convergence and
rounding errors along with other numerical problems, and
all effects of the finite basis and incomplete
correlation treatment is in $\hat H_{rest}$ 
then any conclusions based on the perturbative treatment
therefore does not depend on the size of the
basis set or the level of correlation treatment.

An example of creating several levels of approximate Hamiltonians
can be seen going from the FCI solution in a complete
basis, which we will here take as the exact solution, 
to a truncated CI expansion in a finite basis.
In this case we can write down the CI-matrix or Hamiltonian
in the given basis. The CI-matrix $\bm{H}_0$
\beq
\bm{H}_0 \bm{C} = E \bm{C}
\eeq
multiplied with the CI-vector $\bm{C}$
gives the energy $E$ where the CI-vector $\bm{C}$
\beq
\label{civec}
| \bm{C} \rangle = \sum_i C_i | i \rangle
\eeq
contains the coefficients $C_i$ for the linear
expansion in the Slater determinants $| i \rangle$.

If the Slater determinants in the FCI expansion are contructed directly
from the basis functions and the finite basis set
form a true subset of the complete basis set
then the Slater determinants, in the finite basis, will not change
when the basis is reduced from 
the complete basis to the
finite basis. Reducing to a finite 
basis set will then be 
equivalent to restricting the
number of Slater determinants 
in Eq. \ref{civec} to a finite number $m$, where
all determinants only containing the basis functions of the
finite basis have been included,
\beq
\label{civec2}
| \bm{C}' \rangle = \sum_i^m C_i' | i \rangle
\eeq
and generating a new set of CI-coefficients $C'$.
The approximate CI-matrix $\bm{H}_0'$ can easily be separated
from the exact CI-matrix $\bm{H}_0$, since matrix multiplication
is distributive,
\beq
\bm{H}_0 = \bm{H}_0' + \bm{H}_0 - \bm{H}_0' = \bm{H}_0' + \bm{H}_{rest}
\eeq
with a remaning part $\bm{H}_{rest}$. $\bm{H}_0'$ is
now FCI in a finite basis. The dimension of
$\bm{H}_0'$ is smaller than that of $\bm{H}_0$ and
therefore $\bm{H}_0'$
will only contain approximations to certain solutions in 
$\bm{H}_0$. Furthermore 
not only will the CI-coefficients change going from a
complete basis to a finite basis but also
the matrix elements in $\bm{H}_0'$ will also differ 
from those in $\bm{H}_0$ and hence $\bm{H}_{rest}$
will therefore also be non-zero in the parts where $\bm{H}_0'$
have been subtracted.

If a suitable orbital rotation of the primitive basis
in Eq. \ref{civec2} is performed then the FCI
solution in the finite basis can be reduced 
to the regular CISD solution in the same basis.
We here note that the FCI solution is invariant
to all orbital rotations but the CISD is not.
Arranging the CI-vector $\bm{C}$ in Eq. \ref{civec2} 
according to the regular CI hierarchy
the FCI CI-matrix, in the rotated basis, can be written
\beq
\bm{H}_0' = \sum_{i,j=0}^N \langle i | \bm{H}_0' | j \rangle
\eeq
where the sum over $i,j$ is over all excitation levels
in the CI hierarchy from zero to $N$.
Truncating the CI-expansion at the CISD level
the Hamiltonian $\bm{H}_0''$ can be written as
\beq
\label{cisd}
\bm{H}_0'' = \sum_{i,j=0}^N \langle i | \bm{H}_0' | j \rangle
           - \sum_{i=0,j=3}^N \langle i | \bm{H}_0' | j \rangle
           - \sum_{i=3,j=0}^N \langle i | \bm{H}_0' | j \rangle
           = \sum_{i,j=0}^2 \langle i | \bm{H}_0' | j \rangle.
\eeq 
Solving the CISD equations with the Hamiltonian in 
Eq. \ref{cisd} will give the exact eigenfunctions and eigenvalues for 
$\bm{H}_0''$ which will approximate some solutions in
$\bm{H}_0$. 

When an external perturbation is applied to
the zeroth order Hamiltonian then it is not 
applied to the exact Hamiltonian $\bm{H}_0$
but to some approximative Hamiltonian $\bm{H}_0'$ or $\bm{H}_0''$.
Since the exact eigenfunctions for $\bm{H}_0'$ or $\bm{H}_0''$
are trivially known the discussion about having
exact eigenfunctions and eigenvalues for the zeroth order Hamiltonian
in perturbation theory is redundant. 

\subsection{The length and velocity gauge}
\label{rep}

When approximating $\hat H_0$ by $\hat H_0'$ 
the gauge invariance may be affected since commutation relations
with $\hat H_0'$ may be slightly different for those for $\hat H_0$.
Since the implementation of the quadrupole intensities used here 
is in length gauge \cite{oiqm} 
the conversion from the velocity 
to the length gauge, where $\hat p$ is substituted with $r$,
is central in showing 
how origin independence of the quadrupole intensities
does not hold, in usual quantum chemistry calculations,
in both gauges when $\hat H_0$ is approximated by $\hat H_0'$

Using the Hamiltonian for the Schr{\"o}dinger equation of a
molecular system in the Born-Oppenheimer approximation
in Eq. \ref{tindep} the following commutation
relations are known \cite{bernadotte2012origin} 
\beq
\label{comd}
[ r_{i,\alpha}, \hat H_0 ] = \frac{\imath \hbar}{m} \hat p_{i,\alpha},
\eeq
\beq
\label{comq}
[ r_{i,\alpha} r_{i,\beta} , \hat H_0 ] = 
\frac{\imath \hbar}{m} (\hat p_{i,\alpha} r_{i,\beta} + r_{i,\alpha} \hat p_{i,\beta})
\eeq
\beq
\label{como}
[ r_{i,\alpha} r_{i,\beta} r_{i,\gamma} , \hat H_0 ] = 
\frac{\imath \hbar}{m} (
\hat p_{i,\alpha} r_{i,\beta} r_{i,\gamma} + 
r_{i,\alpha} \hat p_{i,\beta} r_{i,\gamma} +
r_{i,\alpha} r_{i,\beta} \hat p_{i,\gamma} )
\eeq
while other choices of Hamiltonian may not show the
same commutation relations. By using the
commutations relations the different electric terms in the multipole
expansion can be converted from the velocity
to the length representation
\beqa
\langle 0 | \hat \mu_{\alpha}^p | n \rangle &=&
- \imath \frac{E_{0n}}{\hbar}
\langle 0 | \hat \mu_{\alpha} | n \rangle \\
\langle 0 | \hat Q_{\alpha \beta}^p | n \rangle &=&
- \imath \frac{E_{0n}}{\hbar}
\langle 0 | \hat Q_{\alpha \beta} | n \rangle \\
\langle 0 | \hat O_{\alpha \alpha \beta}^p | n \rangle &=&
- \imath \frac{E_{0n}}{\hbar}
\langle 0 | \hat O_{\alpha \alpha \beta} | n \rangle .
\eeqa
The requirement for exact conversion from the velocity to the
length representation is usually stated as having the
exact eigenfunctions for $\hat H_0$. In 
Sec. \ref{approx} it was demonstrated that obtaining
the exact zeroth eigenfunctions is trivial. This, however,
does not mean that the conversion from the velocity
to the length representation is always exact. Since
$\hat H_0$ in Eq. \ref{tindep} has never be solved exactly for $N>2$,
for which the commutations relations in Eqs. \ref{comd}-\ref{como}
is based upon, but only approximate
Hamiltonians $\hat H_0',\hat H_0'', \ldots$
of $\hat H_0$ have been solved. This means
that the commutation relations in Eqs. \ref{comd}-\ref{como}
should not be based on $\hat H_0$ but on an
approximative Hamiltonian $\hat H_0'$
\beq
\label{comda}
[ r_{i,\alpha}, \hat H_0' ] \cong \frac{\imath \hbar}{m} \hat p_{i,\alpha},
\eeq
\beq
\label{comqa}
[ r_{i,\alpha} r_{i,\beta} , \hat H_0' ] \cong 
\frac{\imath \hbar}{m} (\hat p_{i,\alpha} r_{i,\beta} + r_{i,\alpha} \hat p_{i,\beta})
\eeq
\beq
\label{comoa}
[ r_{i,\alpha} r_{i,\beta} r_{i,\gamma} , \hat H_0' ] \cong
\frac{\imath \hbar}{m} (
\hat p_{i,\alpha} r_{i,\beta} r_{i,\gamma} + 
r_{i,\alpha} \hat p_{i,\beta} r_{i,\gamma} +
r_{i,\alpha} r_{i,\beta} \hat p_{i,\gamma} ).
\eeq
The commutation relations in Eqs. \ref{comda}-\ref{comoa}
shows that the conversion from the velocity to the
length representation depends on the 
commutation relations in Eqs. \ref{comda}-\ref{comoa}
and not on having the exact eigenfunctions and eigenvalues for $\hat H_0'$.
If the commutation relations in Eqs. \ref{comda}-\ref{comoa}
become exact the conversion from the velocity to the
length representation exact otherwise it will
only be approximate which we in Sec. \ref{SEC:appl} will use
to demonstrate numerically that exact origin independence is only
found in the velocity gauge in the approximate calculations
performed in quantum chemistry. Even if the commutation
relations in Eqs. \ref{comda}-\ref{comoa} are exact
there is, however, no guarantee that $\hat H_0'$ will be equal
to $\hat H_0$ \cite{olsen_1985}.

\subsection{Origin independence of the oscillator strengths}
\label{perturb}

We will in this section try to recapitulate the ideas and derivations
of Bernadotte \etal \cite{bernadotte2012origin} to show how
$\hat H_0'$ enters and the effects of this along with the points
illustrated in the applications in Sec. \ref{SEC:appl}. For complete
derivations of this topic we refer to Bernadotte \etal \cite{bernadotte2012origin}.

It is thoughout assumed that the electromagnetic fields are weak
and can be treated as a perturbation of the molecular system
which in our case is described by the Schr\"odinger equation 
within the Born-Oppenheimer approximation
\beq
\label{tindep}
\hat H_0 = \sum_{i=1}^{N} \frac{\bm{\hat p}_i^2}{2 m_e} + V(\bm{r}_1, \ldots , \bm{r}_N)
\eeq
where $\hat U(t)$ is the time-dependent perturbation
\beq
\label{tinde}
\hat U(t) = \frac{e A_0}{2 m_e c} \sum_i 
exp(\imath (\bm{k \cdot r}_i -\omega t))( \bm{\mathcal{E} \cdot \hat p}_i) 
\eeq
from a monochromatic linearly polarized electromagnetic wave.
In Eq. \ref{tinde} $\bm{k}$ is the wave vector pointing in the direction of 
propagation, $\bm{\mathcal{E}}$ the polarization vector
perpendicular to $\bm{k}$ and $\omega$ is
the angular frequency. 

By applying Fermi's golden rule and assuming that transitions
only occur when the energy difference between the eigenstates of the unperturbed molecule
matches the frequency of the perturbation
\beq
\omega = \omega_{0n} = \frac{E_n - E_0}{\hbar}
\eeq
the explicit time dependence can be eliminated from the transition rate
\beq
\label{trarate}
\Gamma_{0n}(\omega) = \frac{2 \pi }{\hbar} 
| \langle 0 | \hat U | n \rangle |^2 \delta(\omega-\omega_{0n}) =
\frac{ \pi A_0^2}{2 \hbar c} 
| T_{0n} |^2 \delta(\omega-\omega_{0n}).
\eeq
Where in Eq. \ref{trarate} the transition moments $T_{0n}$ have 
been introduced. The effect of the weak electromagnetic field can now
be expressed as a time-independent expectation value.

In the derivation of Fermi's golden rule only the knowledge
of the exact eigenfunctions and eigenvalues of 
$\hat H_0$ are required. There is no requirement
that a specific $\hat H_0$ must be used
nor does the result depend on any intrinsic properties of
$\hat H_0$. Because of this will any equations derived using
$\hat H_0$ from Eq. \ref{hexact} or $\hat H_0'$ from Eq. \ref{hprime}
only differ in the eigenfunctions and eigenvalues used
and therefore any conclusions, like origin independence,
will also be valid for $\hat H_0'$, irrespectively of
the choice of basis set and level of correlation.

This can also be demonstrated numerically
since exact origin dependence, in the velocity
representation, should only be observed for FCI
in a complete basis if the exact eigenfunctions for
$\bm{H}_0$ in Eq. \ref{tindep} was required and
deviations from exact origin dependence should
be observed in approximate calculations.
We, as expected, always see exact origin dependence
regardless of basis and correlation level in the velocity gauge.
Several numerical examples of the exact origin
dependence will be given in Sec. \ref{SEC:appl} for [FeCl$_4$]$^{1-}$
in different basis sets at the RASSCF level of correlation.

Bernadotte \etal \cite{bernadotte2012origin} showed that origin
independence in the oscillator strengths $f_{0n}$
\beq
\label{oscillator}
f_{0n} = \frac{2 m_e}{e^2 E_{0n}} | T_{0n}|^2 ,
\eeq
where $E_{0n} = E_n - E_0$ is the difference in the
eigenstates of the unperturbed molecule,
comes naturally provided that the collection of the terms in 
Taylor expansion of the exponential of the 
wave vector {\boldmath{$k$}} in Eq. \ref{tinde}
is collected to the same order in the observable oscillator strengths
in Eq. \ref{oscillator} 
\beq
\label{osstr}
f_{0n} = f_{0n}^{(0)} + f_{0n}^{(1)} + f_{0n}^{(2)} + \ldots = \frac{2 m_e}{e^2 E_{0n}} |T_{0n}^{(0)} + T_{0n}^{(1)} + T_{0n}^{(2)} + \ldots |^2
\eeq
and not in the transition moments $T_{0n}$
traditionally done. 

\subsubsection{Isotropically averaged oscillator strengths}

Truncating the expansion of the oscillator strengths
in Eq. \ref{osstr} at the second order gives
the dipole and the quadrupole intensities. 
The zeroth order in Eq. \ref{osstr} is 
the electric-dipole-electric-dipole $f_{0n}^{(\mu^2)}$ contribution
\beq
\langle f_{0n}^{(\mu^2)} \rangle _{iso} = 
\frac{2 m_e}{3 e^2 \hbar ^2} E_{0n} 
\sum_{\alpha} \langle 0 | \hat \mu_{\alpha} | n \rangle^2 =
\frac{2 m_e}{3 e^2 \hbar ^2} E_{0n} 
\langle 0 | \bm{ \hat \mu} | n \rangle^2
\eeq
where the sum is over $x,y,z$ if Cartesian coordinates
is used. The first order $f_{0n}^{(1)}$ in Eq. \ref{osstr} 
vanishes while the second order $f_{0n}^{(2)}$ gives four
non-zero contributions. The electric-quadrupole-electric-quadrupole $ f_{0n}^{(Q^2)}$
\beq
\label{eqeq}
\langle f_{0n}^{(Q^2)} \rangle _{iso} =
\frac{m_e}{20 e^2 \hbar^4 c^2} E_{0n}^3
\left[ \sum_{\alpha \beta} \langle 0 | \hat Q_{\alpha \beta} | n \rangle^2 -
\frac{1}{3} ( \sum_{\alpha} \langle 0 | \hat Q_{\alpha \alpha} | n \rangle)^2 \right],
\eeq
the magnetic-dipole-magnetic-dipole $f_{0n}^{(m^2)}$
\beq
\label{mdmd}
\langle f_{0n}^{(m^2)} \rangle _{iso} =
\frac{2 m_e}{3 e^2 \hbar ^2} E_{0n}
\sum_{\alpha} \langle 0 | \hat m_{\alpha} | n \rangle^2 =
\frac{2 m_e}{3 e^2 \hbar ^2} E_{0n}
\langle 0 | \bm{ \hat m} | n \rangle^2 ,
\eeq
the electric-dipole-electric-octupole $f_{0n}^{(\mu O)}$
\beq
\label{edeo}
\langle f_{0n}^{(\mu O)} \rangle _{iso} =
-\frac{2 m_e}{45 e^2 \hbar^4 c^2} E_{0n}^3
\sum_{\alpha \beta} \langle 0 | \hat \mu_{\beta} | n \rangle \langle 0 | \hat O_{\alpha \alpha \beta} | n \rangle ,
\eeq
and the electric-dipole-magnetic-quadrupole $f_{0n}^{(\mu \mathcal{M})}$ contributions
\beq
\label{edmq}
\langle f_{0n}^{(\mu \mathcal{M})} \rangle _{iso} = 
\frac{m_e}{3 e^2 \hbar^3 c} E_{0n}^2
\sum_{\alpha \beta \gamma} \varepsilon_{\alpha \beta \gamma}
\langle 0 | \hat \mu_{\beta} | n \rangle Im \langle 0 |  \mathcal{\hat M} _{\gamma \alpha} | n \rangle
\eeq
which all have to be included to obtain origin independence.

\subsubsection{Origin dependence of the transition moments}
\label{odtm}

As shown in \cite{bernadotte2012origin} the individual terms
in the expansion of the oscillator strengths in Eqs. \ref{eqeq}-\ref{edmq} are not
individually origin independent
but rely on exact cancellation for the total oscillator strength order by order.
The proof of the exact cancellation after the multipole expansion
is more complicated than that for the exact expression
repeated in Appendix \ref{app:exact} and we will therefore refer to
Bernadotte \etal \cite{bernadotte2012origin} for the proof.
When the origin is shifted from $\bm{O}$ to $\bm{O} + \bm{a}$, in the velocity gauge,
the electric-quadrupole transition moments
\beq
\label{eqm}
\langle 0 | \hat Q_{\alpha \beta}^p(\bm{O} + \bm{a}) | n \rangle =
\langle 0 | \hat Q_{\alpha \beta}^p(\bm{O} ) | n \rangle -
a_{\beta} \langle 0 | \hat \mu_{\alpha}^p | n \rangle -
a_{\alpha} \langle 0 | \hat \mu_{\beta}^p | n \rangle,
\eeq
where the $\alpha, \beta$ are the different $x,y,z$ components,
the electric-octupole transition moments
\beqa
\label{eom}
\langle 0 | \hat O_{\alpha \alpha \beta}^p(\bm{O} + \bm{a}) | n \rangle &=&
\langle 0 | \hat O_{\alpha \alpha \beta}^p(\bm{O} ) | n \rangle \nonumber \\ &-&
a_{\gamma} \langle 0 | \hat Q_{\alpha \beta}^p(\bm{O} ) | n \rangle -
a_{\beta} \langle 0 | \hat Q_{\alpha \gamma}^p(\bm{O} ) | n \rangle -
a_{\alpha} \langle 0 | \hat Q_{\beta \gamma}^p(\bm{O} ) | n \rangle \nonumber \\ &+&
a_{\alpha}a_{\beta} \langle 0 | \hat \mu_{\gamma}^p | n \rangle +
a_{\alpha}a_{\gamma} \langle 0 | \hat \mu_{\beta}^p | n \rangle +
a_{\beta}a_{\gamma} \langle 0 | \hat \mu_{\alpha}^p | n \rangle,
\eeqa
the magnetic-dipole transition moments
\beqa
\label{mdm}
\langle 0 | \hat m_{\alpha}(\bm{O} + \bm{a}) | n \rangle &=&
\langle 0 | \hat m_{\alpha}(\bm{O} ) | n \rangle -
\varepsilon_{\alpha \beta \gamma} a_{\beta} \frac{1}{2c} \langle 0 | \hat \mu_{\gamma}^p | n \rangle \nonumber \\ &\cong&
\langle 0 | \hat m_{\alpha}(\bm{O} ) | n \rangle -
\varepsilon_{\alpha \beta \gamma} a_{\beta} \frac{\imath E_{0n}}{2 \hbar c} \langle 0 | \hat \mu_{\gamma} | n \rangle,
\eeqa
where $\varepsilon_{\alpha \beta \gamma}$ is the Levi-Civita tensor,
and the magnetic-quadrupole transition moments
\beqa
\label{mqm}
\langle 0 |  \mathcal{\hat M} _{\gamma \alpha}(\bm{O} + \bm{a}) | n \rangle &=&
\langle 0 |  \mathcal{\hat M} _{\gamma \alpha}(\bm{O} ) | n \rangle \nonumber \\ &-&
\frac{1}{3c} \varepsilon_{\alpha \gamma \delta} a_{\gamma} \langle 0 | \hat Q_{\beta \delta}^p(\bm{O} ) | n \rangle +
\frac{2}{3c} \varepsilon_{\alpha \gamma \delta} a_{\beta}a_{\gamma} \langle 0 | \hat \mu_{\delta}^p | n \rangle \nonumber \\ &+&
\frac{2}{3} \delta_{\alpha \beta} (\bm{a} \cdot \langle 0 | \bm{\hat m}(\bm{O} ) | n \rangle -
2 a_{\beta} \langle 0 | \hat m_{\alpha}(\bm{O} ) | n \rangle \nonumber \\ &\cong&
\langle 0 |  \mathcal{\hat M} _{\gamma \alpha}(\bm{O} ) | n \rangle \nonumber \\ &+&
\frac{\imath E_{0n}}{3 \hbar c} \varepsilon_{\alpha \gamma \delta} a_{\gamma} \langle 0 | \hat Q_{\beta \delta} (\bm{O} ) | n \rangle -
\frac{\imath 2 E_{0n}}{3 \hbar c} \varepsilon_{\alpha \gamma \delta} a_{\beta}a_{\gamma} \langle 0 | \hat \mu_{\delta} | n \rangle \nonumber \\ &+&
\frac{2}{3} \delta_{\alpha \beta} (\bm{a} \cdot \langle 0 | \bm{\hat m}(\bm{O} ) | n \rangle -
2 a_{\beta} \langle 0 | \hat m_{\alpha}(\bm{O} ) | n \rangle
\eeqa
all produce all lower order contributions which 
are all in the velocity gauge.
In the magnetic terms in Eqs. \ref{mdm} and \ref{mqm}
the $\mu_{\delta}^p$ and $\hat Q_{\beta \delta}^p$ terms
have been transformed from the velocity to the length gauge
as described in Sec. \ref{rep}. We here note that the
energy appearing in the transformation from the velocity
to the length gauge is calculated exactly for $\hat H_0'$
and the error is therefore only from the commutation relations
in Eqs. \ref{comda}-\ref{comoa}.
Transforming the electric terms in Eqs. \ref{eqm} and \ref{eom}
from the velocity to the length gauge
\beq
\label{eqmp}
\langle 0 | \hat Q_{\alpha \beta}(\bm{O} + \bm{a}) | n \rangle =
\langle 0 | \hat Q_{\alpha \beta}(\bm{O} ) | n \rangle -
a_{\beta} \langle 0 | \hat \mu_{\alpha} | n \rangle -
a_{\alpha} \langle 0 | \hat \mu_{\beta} | n \rangle
\eeq
and 
\beqa
\label{eomp}
\langle 0 | \hat O_{\alpha \alpha \beta}(\bm{O} + \bm{a}) | n \rangle &=&
\langle 0 | \hat O_{\alpha \alpha \beta}(\bm{O} ) | n \rangle \nonumber \\ &-&
a_{\gamma} \langle 0 | \hat Q_{\alpha \beta}(\bm{O} ) | n \rangle -
a_{\beta} \langle 0 | \hat Q_{\alpha \gamma}(\bm{O} ) | n \rangle -
a_{\alpha} \langle 0 | \hat Q_{\beta \gamma}(\bm{O} ) | n \rangle \nonumber \\ &+&
a_{\alpha}a_{\beta} \langle 0 | \hat \mu_{\gamma} | n \rangle +
a_{\alpha}a_{\gamma} \langle 0 | \hat \mu_{\beta} | n \rangle +
a_{\beta}a_{\gamma} \langle 0 | \hat \mu_{\alpha} | n \rangle
\eeqa
only produce lower order terms in the length gauge.

The conversion from velocity to length gauge
is, however, only exact if the commutation 
relations in Sec. \ref{rep} are exact.
If the commutations relations in Sec. \ref{rep} are not exact
the magnetic-dipole and magnetic-quadrupole transition
moments will not show exact origin dependence
in the length representation 
and hence the origin independence
of the total oscillator strengths will not be conserved.
We here note that the magnetic terms in the multipole expansion is
not transformed when going from the velocity to the length gauge. 

In the perturbative
inclusion of the electromagnetic fields it is, as always,
asumed that the exact eigenfunctions to $\hat H_0$ is known.
If an approximate wavefunction to $\hat H_0$ is used 
the origin dependence of the various second order oscillator strengths
contributions in Eqs. \ref{eqm}-\ref{mqm} need no longer be exact and hence
origin independence is no longer guaranteed
since the origin independence of the oscillator 
strengths rely on exact cancellation. With the 
above interpretation, which is the prevalent
interpretation, the perturbative requirement is
therefore only exactly fulfilled for full configuration interaction (FCI)
in a complete basis.

We will, however, show that the the requirement for 
having the exact eigenfunctions is trivially fulfilled
and using the velocity representation will automatically
insure origin independence, if all terms for a given order
in the oscillator strengths $f_{0n}$ in Eq. \ref{osstr} is kept,
while using the length
representation will depend on how exact the 
commutation relations in Eqs. \ref{comda}-\ref{comoa} are.

\section{Application}
\label{SEC:appl}

To numerically prove the exact origin independence 
in the velocity gauge for any basis set or level of correlation
we will use the recently
implemented origin independent quadrupole intensites \cite{oiqm}
part in MOLCAS \cite{molcas8}. Since our implementation
is in the length gauge and $f_{0n}^{(\mu^2)}$ is the only electric term
implemented in both the velocity and length gauge
our implementation does not show exact origin independence
unlike those where the velocity gauge is used \cite{bernadotte2012origin,list2015beyond,list2016average,Lestrange}.
However, since the origin independence relies on exact cancellation it is
sufficient to show the exact origin dependence of the
different terms in Eqs. \ref{eqm}-\ref{mqm} in the 
velocity gauge for the [FeCl$_4$]$^{1-}$ molecule
using different basis sets and level of correlation.
In the length gauge the exact origin 
dependence is only found for the electric terms
and not for the magnetic terms as can be seen
from Eqs. \ref{mdm}-\ref{eomp}.

\subsection{Computational details}

We have choosen the [FeCl$_4$]$^{1-}$ molecule
due to its significant increase in pre-edge intensity,
through 4p mixing, in X-ray absorption spectroscopy (XAS) \cite{westre1997multiplet,guo_2016}. 
The 4p mixing gives rise to very large 
$f_{0n}^{(\mu^2)}$ and hence makes the terms in 
Eqs. \ref{eqm}-\ref{eomp} grow significantly faster
and thereby making the conservation
of origin independence more difficult \cite{oiqm}.

We have thoughout used the ANO-RCC basis sets since these
basis sets have been shown to perform reasonanbly
well in conserving the origin independence in 
the length representation for the quadrupole intensities \cite{oiqm}.
Furthermore we have included AUG-cc-pVDZ basis set,
which in a previous application on [FeCl$_4$]$^{1-}$
gave unphysical results, to show that good
basis sets are not needed to have exact
origin independence in the velocity gauge but that
the length gauge is very sensitive to this.

For the correlation treatment all calculations
will be at the RASSCF level, since the FCI limit
cannot be reached, with 
the 1s core electrons in RAS1 and  
11 electrons in 13 orbitals in RAS2. Here
[FeCl$_4$]$^{1-}$ will have $T_d$ geometry
with an Fe-Cl distance of 2.186 {\AA} and the
orbitals for the core-excited states will 
be averaged over 70 states. The intensities are
calculated using the RASSI program \cite{malmqvist1989casscf,rassi_so}
which uses a biorthonormalization procedure
which removes the gauge dependence of non-orthorgonal states.

While the Hamiltonian used in the derivation of the
intensities in Sec. \ref{perturb} is based on the
Schr{\"o}dinger equation we will use a second-order
Douglas-Kroll-Hess Hamiltonian \cite{douglas,Hess:1986} to take into 
account the scalar relativistic effects,
however, as shown in Sec. \ref{approx} the choice of $\hat H_0$
does not matter.

While it would be sufficient only to run two calculations
with different origins to show that the quadrupole intensities
are origin independent in the velocity and not in the
length gauge we will try to vary the basis set
to illustrate the point
that the perturbation does depend on the choice of $\hat H_0$,
as stated in Sec. \ref{approx}, and that the in the length
gauge the electric terms will show exact origin dependence
but the magnetic terms will not. We will therefore show
the exact origin dependence of the electric terms in the length
gauge along with the exact origin dependence of $f_{0n}^{(m^2)}$
in the velocity gauge and the basis set dependence in the length
gauge.

\subsection{Electric terms}

As shown in Eq. \ref{eqmp} the origin dependence of the 
electric-quadrupole transition moments in the length gauge
is exact, up to numerical rounding, in the approximate
calculations performed in electronic structure theory.
In Figure \ref{figeqm} $ f_{0n}^{(Q^2)}$ for the third 
core excited state in the length
gauge in different basis sets have been plotted.
$ f_{0n}^{(Q^2)}$ is seen
to increase rapidly as the origin is moved in the $Z$-direction which is
due to the very large $f_{0n}^{(\mu^2)}$, compared to $ f_{0n}^{(Q^2)}$, 
as can be seen in Table \ref{fecl4basis}. The 
error curves in Figure \ref{figeqm} show the difference
between moving the origin and calculating the 
effect of moving the origin from Eq. \ref{eqmp}. Since
single precision have been used the difference is in the
8th digit and occasional in the 7th as would be expected
due to numerical noise from the finite accuracy and the
origin dependence in the length gauge for $ f_{0n}^{(Q^2)}$
is therefore exact when disregarding numerical noise.

\bef[h!]
\bc
\includegraphics[width=0.5\textwidth,angle=270]{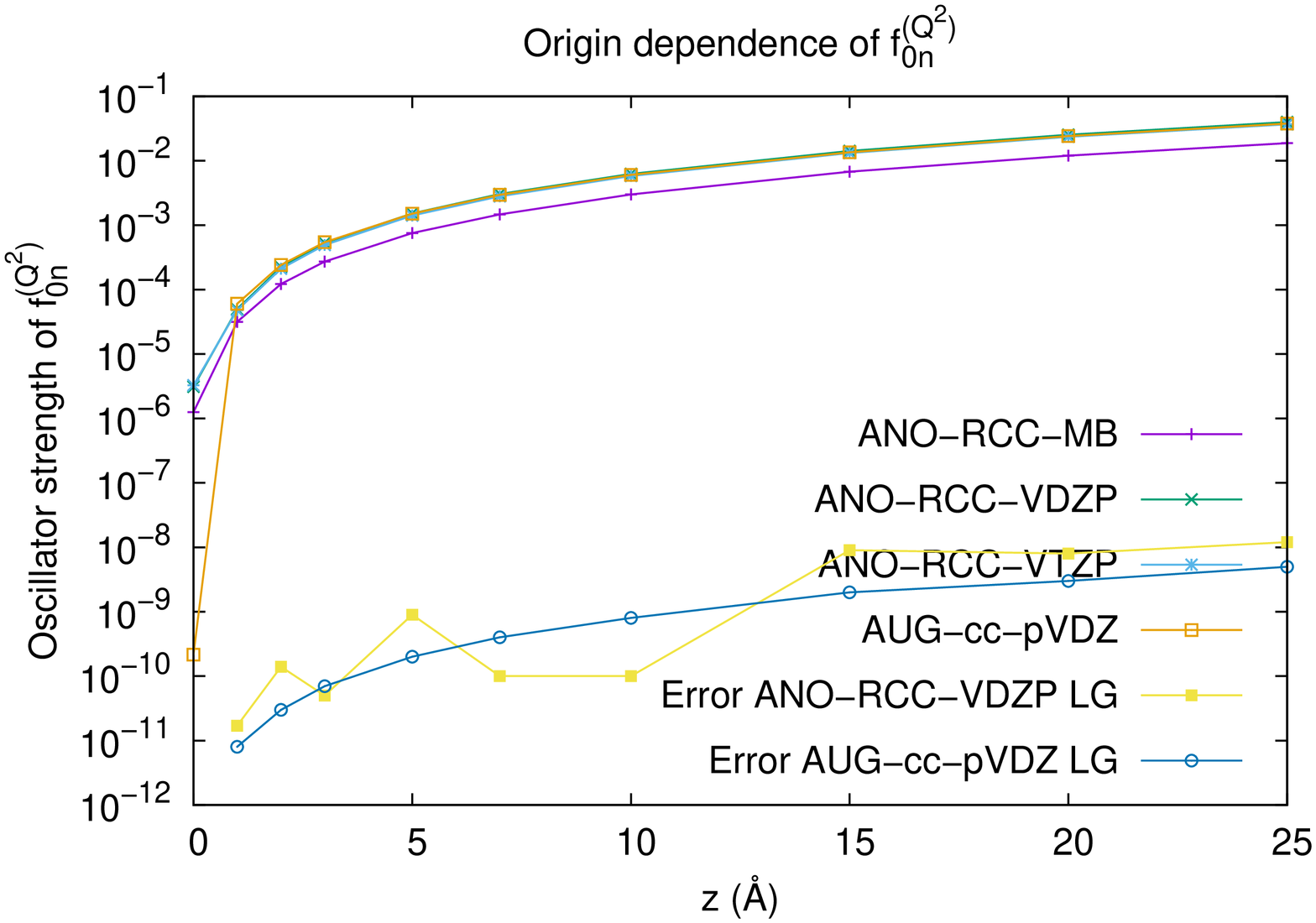}
\caption{The origin dependence of $ f_{0n}^{(Q^2)}$ in different basis sets.
         The error curves shows the numerical error in the origin dependence
         in the given basis set in the length gauge (LG) caused by numerical noise. }
\label{figeqm}
\ec
\ef 

Despite the fact that the AUG-cc-pVDZ underestimates the
$ f_{0n}^{(Q^2)}$ contribution by four magnitudes the origin
dependence of $ f_{0n}^{(Q^2)}$ is very well behaved which
numerically demonstrates that the origin dependence is
independent of the quality of the basis set.

\bt
\btb{lcccccc}
Basis & $f_{0n}^{(\mu^2)}$ & $f_{0n}^{(\mu^2)^p}$ & $R_{dip}$ & $ f_{0n}^{(Q^2)}$ & $f_{0n}^{(\mu O)}$ & $f_{0n}^{(m^2)}$ \\ \hline
ANO-RCC-MB   & 0.115 & 0.111 & 1.04 & 0.0125 & 0.00588 & $0.347*10^{-18}$ \\
ANO-RCC-VDZP & 0.295 & 0.286 & 1.03 & 0.0309 & 0.0208 &  \\
ANO-RCC-VTZP & 0.283 & 0.273 & 1.03 & 0.0327 & 0.0196 &  \\
AUG-cc-pVDZ  & 0.281 & 0.168 & 1.68 & $0.215*10^{-5}$ & 0.583 & $0.341*10^{-11}$
\etb
\caption{The electric-dipole-electric-dipole ($f_{0n}^{(\mu^2)}$), in the length and velocity gauge,
         electric-quadrupole-electric-quadrupole ($f_{0n}^{(Q^2)}$),
         electric-dipole-electric-octupole ($f_{0n}^{(\mu O)}$), both in the length gauge, and
         magnetic-dipole-magnetic-dipole $f_{0n}^{(m^2)}$, same in both gauges, 
         intensities for the
         transition from the ground state to the third 
         core-excited state in [FeCl$_4$]$^{1-}$ in different basis sets
         along with the ratio between the dipole intensities $R_{dip}$.
         All values have been multiplied by $10^{4}$ and values below $10^{-19}$ have been omitted.} 
\label{fecl4basis}
\et

The $f_{0n}^{(\mu O)}$ contribution also shows exact origin dependence,
up to numerical rounding, for all basis sets as shown in Fig. \ref{figmuo}.
All contributions from $f_{0n}^{(\mu O)}$ are negative and gives a contribution
that is only slightly smaller than $ f_{0n}^{(Q^2)}$ in the ANO-RCC basis sets.
In the AUG-cc-pVDZ basis set the $f_{0n}^{(\mu O)}$ contribution is very large
which gives a total negative intensity for this transition as also reported
earlier \cite{oiqm,Lestrange}. By including the fourth order in the intensity the $f_{0n}^{(OO)}$ term
should retify the problem of total negative intensities 
provided that no other higher terms also grows disproportionately
large. The convergence behaviour of the multipole expansion is, however,
not obvious.

\bef[h!]
\bc
\includegraphics[width=0.5\textwidth,angle=270]{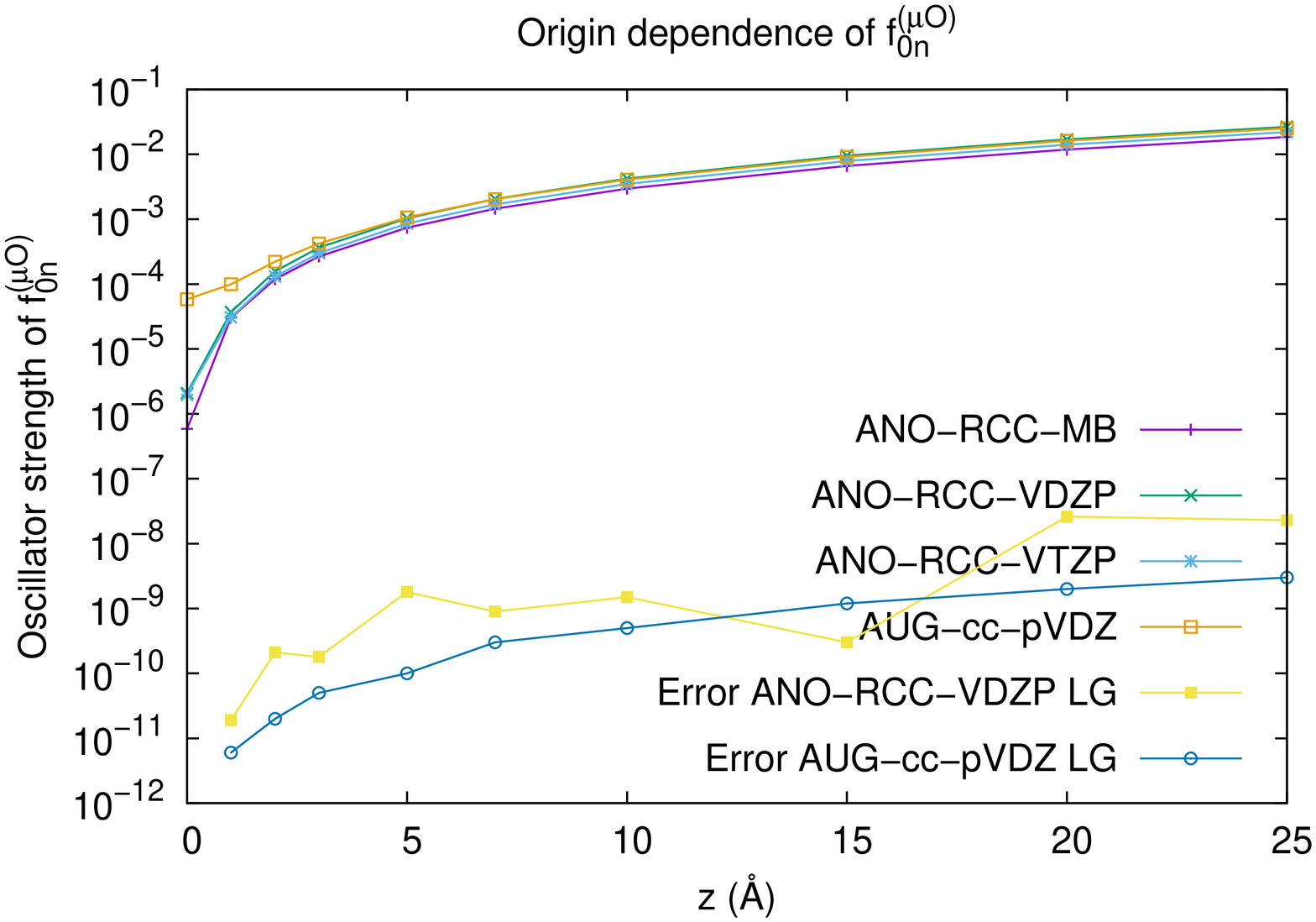}
\caption{The origin dependence of $f_{0n}^{(\mu O)}$ in different basis sets.
         The error curves shows the numerical error in the origin dependence
         in the given basis set in the length gauge (LG) caused by numerical noise.}
\label{figmuo}
\ec
\ef 

\subsection{Magnetic terms}

While all the electric terms shows exact origin dependence in both
the velocity and length gauge the same is not true for the magnetic
terms. Since the magnetic terms are not transformed when changing from the
velocity to the length gauge the displacement of the origin will
therefore depend on $f_{0n}^{(\mu^2)^p}$ in both gauges. Hence in the
length gauge this will introduce an error that will depend on the
difference between $f_{0n}^{(\mu^2)^p}$ and $f_{0n}^{(\mu^2)}$
\beq
\label{prdip}
\Delta =  f_{0n}^{(\mu^2)^p} - f_{0n}^{(\mu^2)}
       =  f_{0n}^{(\mu^2)^p} ( 1 - \frac{f_{0n}^{(\mu^2)}}{f_{0n}^{(\mu^2)^p}} )
       =  f_{0n}^{(\mu^2)^p} ( 1 - R_{dip})
\eeq
where the severity of the error will depend on the 
size of $f_{0n}^{(\mu^2)^p}$ and the ratio $R_{dip}$, both
shown in Table \ref{fecl4basis}.
For $f_{0n}^{(m^2)}$ the dependence on $\Delta$ will be quadratic as
can be seen from Eq. \ref{mdmd} and Eq. \ref{mdm}. 

\bef[h!]
\bc
\includegraphics[width=0.5\textwidth,angle=270]{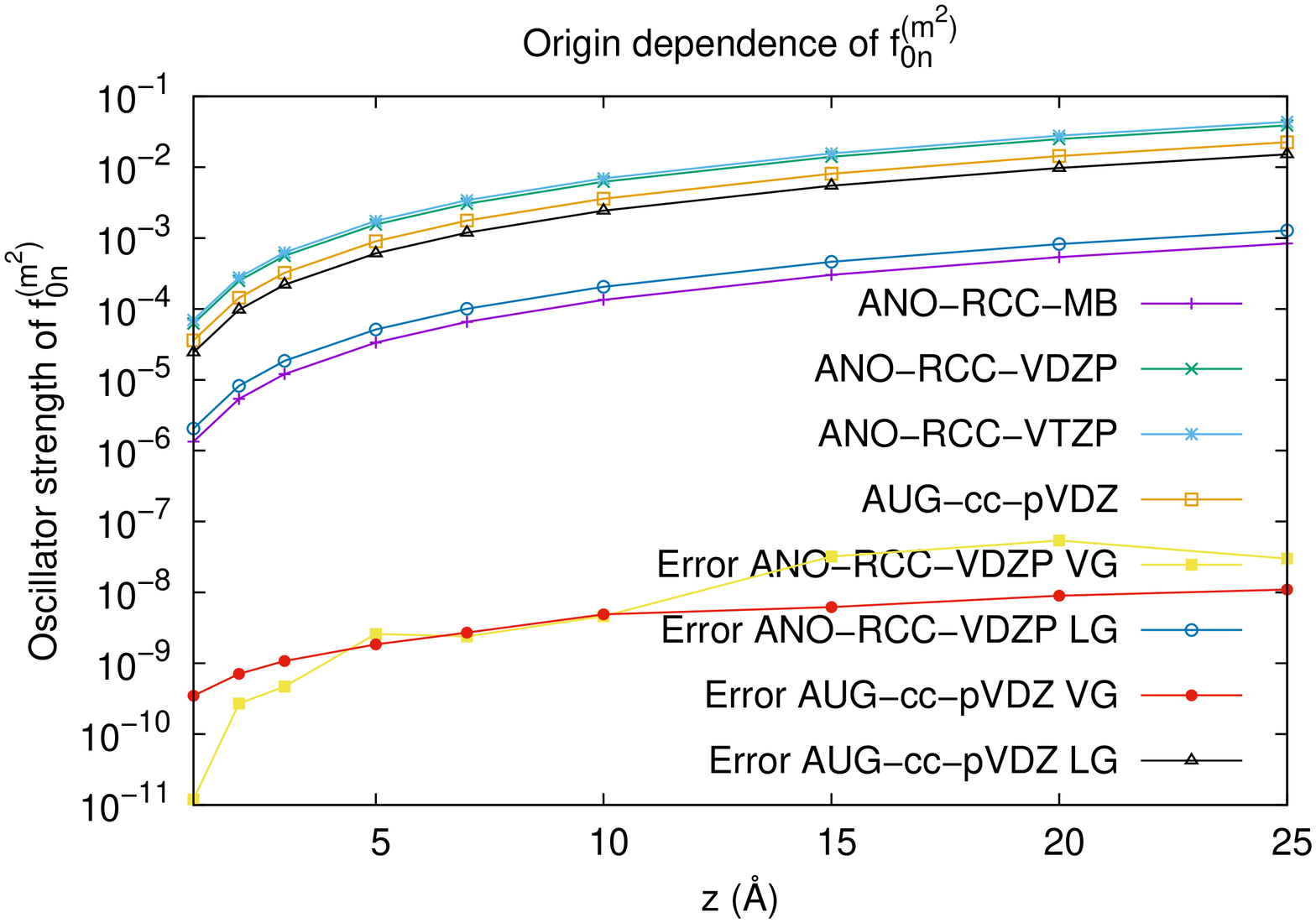}
\caption{The origin dependence of $f_{0n}^{(m^2)}$ in both the velocity
         and length gauge in different basis sets. 
         The error curves shows the numerical error in the origin dependence
         in the given basis set in the length gauge (LG) and velocity gauge (VG).}
\label{figmd}
\ec
\ef 

Fig. \ref{figmd} shows the origin dependence of the $f_{0n}^{(m^2)}$
contribution. The point where the origin of the coordinate system
coincides with the Fe atom have been omitted to better show the 
origin dependence since the $f_{0n}^{(m^2)}$
contribution is negligible when the origin is placed on the 
Fe atom, as can be seen in Table \ref{fecl4basis}.

In the velocity gauge the origin dependence of $f_{0n}^{(m^2)}$
is exact, down to numerical noise, as can be seen from the error
curves labelled with VG in Fig. \ref{figmd}. In the length
gauge, however, there is a strong dependence on the origin
and basis set as can be seen from by comparing the error curves
in the AUG-cc-pVDZ and ANO-RCC-VDZP basis sets labelled with LG.
In the ANO-RCC-VDZP error curve in the length gauge the
difference is two orders of magnitude smaller than $f_{0n}^{(m^2)}$
while in the AUG-cc-pVDZ basis set the error is almost the same size
as $f_{0n}^{(m^2)}$ which shows that if the origin is placed
close to the Fe atom, less than 3 \AA, the ANO-RCC-VDZP will produce reliable results
while AUG-cc-pVDZ basis set cannot.

\section{Conclusion}
\label{SEC:summ}

We have here discussed the consequences of not having the
exact eigenfunctions and eigenvalues for $\hat H_0$ in perturbation
theory. The usual approximations such as projecting
the wave function on to a finite basis set and restricting
the particle interaction usually used for finding the
eigenfunctions and eigenvalues for $\hat H_0$ 
is in fact a way of constructing an approximate or effective zeroth order 
Hamiltonian $\hat H_0'$. 
It it here shown that if the perturbation expansion
does not depend on any intrinsic properties of $\hat H_0$ 
but only rely on $\hat H_0$ having a spectrum then any
$\hat H_0'$, which also have a spectrum,
will also give the exact same perturbation expansion.
Any conclusion or statement reached from the 
perturbation expansion for $\hat H_0$ will therefore also 
be valid for $\hat H_0'$. Since $\hat H_0'$ per
definition is always solved exactly the 
exact eigenfunctions and eigenvalues for $\hat H_0'$
is always known and since $\hat H_0'$ 
is the zeroth order Hamiltonian used the question
about having the exact eigenfunctions and eigenvalues
for the zeroth order Hamiltonian is redundant since
this is trivially fulfilled for $\hat H_0'$.

Since Fermi's golden rule, which only require that
$\hat H_0$ has a spectrum, is used in the derivation
of the origin independent intensities \cite{bernadotte2012origin}
it is therefore trival to show that this will
hold for any approximative $\hat H_0'$ which 
also have a spectrum. The origin independence of the intensities
in the velocity gauge therefore always hold
irrespectively of the choice of basis set
and level of correlation as also demonstrated numerically.
In the length gauge the origin independence
is, however, not gauranteed and only rely
on how well $\hat H_0'$ reproduce the commutation 
relations of $\hat H_0$ since the
exact eigenfunctions and eigenvalues of $\hat H_0'$
is always known.

The calculation of the intensities presented here can be performed
significantly more elegant  
by calculating the exact expression, from Appendix \ref{app:exact},
as shown by List \etal \cite{list2015beyond,list2016average}.
Here the multipole expansion is completely avoided
and origin independence is also a given in the
velocity gauge, as also discussed in Appendix \ref{app:exact}
and \cite{list2015beyond}.

\section*{Acknowledgement(s)}
Financial support was received
from the Knut and Alice Wallenberg Foundation for the
project “Strong Field Physics and New States of Matter”
(Grant No. KAW-2013.0020) and the Swedish Research Council (Grant No. 2012-3910).
Computer resources were provided by SNIC trough the National Supercomputer Centre at Link{\"o}ping University (Triolith) under projects snic2014-5-36,  snic2015-4-71, snic2015-1-465 and snic2015-1-427.


\begin{appendix}
 \section{The exact expressions}
\label{app:exact}

Showing the origin independence for the intensities
for the exact expression is well known and
significantly easier to show than
the origin independence for the intensities
after the multipole expansion \cite{bernadotte2012origin}.
We will here repeat the derivation for the exact expression for the the sake
of completeness and to show that this is 
connected to the discussion in Sections \ref{sec:pert}
and \ref{approx}. The exact expression in the velocity
gauge
\beqa
| T_{0n}(\bm{O} + \bm{a})|^2 &=& 
              \langle 0 | exp( \imath \bm{k \cdot (r-a)})( \bm{\mathcal{E} \cdot \hat p}) | n \rangle
              \langle n | exp(-\imath \bm{k \cdot (r-a)})( \bm{\mathcal{E} \cdot \hat p}) | 0 \rangle \nonumber \\
                             &=& 
              \langle 0 | exp( \imath \bm{k \cdot r})( \bm{\mathcal{E} \cdot \hat p}) | n \rangle
              \langle n | exp(-\imath \bm{k \cdot r})( \bm{\mathcal{E} \cdot \hat p}) | 0 \rangle exp( \imath \bm{k \cdot (a-a)}) \nonumber \\
                             &=&
              \langle 0 | exp( \imath \bm{k \cdot r})( \bm{\mathcal{E} \cdot \hat p}) | n \rangle
              \langle n | exp(-\imath \bm{k \cdot r})( \bm{\mathcal{E} \cdot \hat p}) | 0 \rangle \nonumber \\
                             &=& | T_{0n}(\bm{O})|^2
\eeqa
will show exact origin independence provided that the $\hat H_0'$
used has a spectrum. We here note that $| 0 \rangle$ and $| n \rangle$
are exact eigenfunctions of $\hat H_0'$ and that there is no requirement
of the choice of basis or level or correlation. 

Using Eq. \ref{comda} to transform from the velocity
to the length gauge
\beqa
| T_{0n}(\bm{O} + \bm{a})|^2 \cong \frac{-m^2}{\hbar^2}
      &\langle& 0 | exp( \imath \bm{k \cdot (r-a)})( \bm{\mathcal{E} \cdot ((r-a) \hat H_0' - \hat H_0' (r-a))} ) | n \rangle \times \nonumber \\
      &\langle& n | exp(-\imath \bm{k \cdot (r-a)})( \bm{\mathcal{E} \cdot ((r-a) \hat H_0' - \hat H_0' (r-a))} ) | 0 \rangle \nonumber \\
                             \cong \frac{-m^2}{\hbar^2}
    ( &\langle& 0 | exp( \imath \bm{k \cdot r})( \bm{\mathcal{E} \cdot (r \hat H_0' - \hat H_0' r)} ) | n \rangle + \nonumber \\
      &\langle& 0 | exp( \imath \bm{k \cdot r})( \bm{\mathcal{E} \cdot (a \hat H_0' - \hat H_0' a)} ) | n \rangle ) \times \nonumber \\
    ( &\langle& n | exp(-\imath \bm{k \cdot r})( \bm{\mathcal{E} \cdot (r \hat H_0' - \hat H_0' r)} ) | 0 \rangle + \nonumber \\
      &\langle& n | exp(-\imath \bm{k \cdot r})( \bm{\mathcal{E} \cdot (a \hat H_0' - \hat H_0' a)} ) | 0 \rangle ) exp( \imath \bm{k \cdot (a-a)})\nonumber \\
                             \cong | T_{0n}(\bm{O}&)&|^2
\eeqa
gives a second order polynomial in $\bm{a}$ where the zeroth order gives non-zero contribution
and all higher orders are zero
since $\hat H_0'$ commutes with $\bm{a}$. If the
commutation relations for $\hat H_0$ and $\hat H_0'$
are the same then the origin independence in the
length gauge will hold for the exact expression.

\end{appendix}

\bibliography{full}

\end{document}